\documentclass[preprint,aps,nofootinbib]{revtex4-1}
\pdfoutput=1

\usepackage{amsmath}  
\usepackage{amsbsy,amssymb,amsfonts}
\usepackage{graphicx}
\usepackage{color}
\usepackage{epsf}
\usepackage[utf8x]{inputenc}
\usepackage{hyperref}
\usepackage{multirow}
\def\ie{{\em{i.e.}},}

\def\bravert{\egroup\,\vrule\,\bgroup}

{\catcode`\|=\active
  \gdef\Twoint#1{\left(\mathcode`\|"8000\let|\bravert {#1}\right)}}
{\catcode`\|=\active
  \gdef\Braket#1{\left<\mathcode`\|"8000\let|\bravert {#1}\right>}}

\newcommand{\beq}{\begin{equation}}
\newcommand{\eeq}{\end{equation}}
\newcommand{\beqa}{\begin{eqnarray}}
\newcommand{\eeqa}{\end{eqnarray}}
\newcommand{\bea}{\begin{array}}
\newcommand{\eea}{\end{array}}

\newcommand{\bef}{\begin{figure}}
\newcommand{\ef}{\end{figure}}
\newcommand{\bc}{\begin{center}}
\newcommand{\ec}{\end{center}}
\newcommand{\bt}{\begin{table}}
\newcommand{\et}{\end{table}}
\newcommand{\btb}{\begin{tabular}}
\newcommand{\etb}{\end{tabular}}

\newcommand{\au}{{\em a.u.}}

\def\rvac{\left| \rule{0.3cm}{.0cm} \right>}

\def\au{{\it a.u.\ }}
\begin{document}

\title { Atomic Clock States of Thulium have Near-Zero Electric Quadrupole Moment }

\vspace*{1cm}

\author{Timo Fleig}
\email{timo.fleig@irsamc.ups-tlse.fr}
\affiliation{Laboratoire de Chimie et Physique Quantiques,
             FeRMI, Universit{\'e} Paul Sabatier Toulouse III,
             118 Route de Narbonne, 
             F-31062 Toulouse, France }
\vspace*{1cm}
\date{\today}

\vspace*{1cm}
\begin{abstract}
A method for highly accurate calculations of atomic electric quadrupole moments (EQM) is presented, using relativistic 
general-excitation-rank configuration interaction wavefunctions based on Dirac spinors. Application 
to the clock transition states of the thulium atom 
employing up to full Quadruple excitations for the atomic wavefunction
yields a final value of $Q_{zz}({^2F}_{7/2}) = 0.07 \pm 0.07$ \au,
establishing that the thulium electronic ground state has an exceptionally small EQM. A detailed analysis of this
result is presented which has implications for EQMs of other atoms with unpaired $f$ electrons.
\end{abstract}

\maketitle
\section{Introduction}
\label{SEC:INTRO}
In atomic systems electric quadrupole moments of the electron shells are of importance in the quest for setting new 
standards of time measurement. In fact, they are often 
one of the limiting quantities in optical atomic clock's fractional frequency uncertainty \cite{safronova_rmp_2018}. 
The reason for this type of uncertainty is that in many implementations of atomic clocks an environmental perturbation,
a residual external electric-field gradient, interacts with the EQM of the (atomic) charge distribution in the relevant 
clock-transition states, giving rise to the so-called ``quadrupole shifts'' \cite{ludlow_rmp_2015} of atomic energy
levels.

Lanthanide atoms have electronic valence shells $4f^k$ where the $k$ $f$ electrons are partially shielded from external electric
fields by the electrons occupying the $6s$ shell. This makes $f-f$ transitions in lanthanides less susceptible to EQM
interactions with the gradients of an external electric field \cite{Dzuba_Er_2013}. Indeed, great progress has very
recently been made in devising a transportable optical clock using the hyperfine levels of the ground-state $f-f$ transition
in thulium atoms \cite{NatComm2_Tm_2021,MOT_transp_Tm_2021,CloSta_Tm_2020,NatComm_Tm_2019}. The EQM interaction is 
nevertheless one of the sources of uncertainty in this type of atomic clock.
However, electronic properties of lanthanide atoms are notoriously difficult to calculate accurately.

The method of calculating EQMs presented in this paper is based on relativistic configuration interaction (CI) wavefunctions 
of general excitation rank, up to the level of Full CI \cite{knecht_luciparII}. 
The implementation is highly efficient, allowing for CI expansions 
with up to $10^{10}$ ($10$ billion) linear expansion terms and thus for very accurate calculation of electron correlation 
effects. This will be demonstrated in the present paper.
Moreover, the present approach is particularly advantageous when atomic states with complicated
shell structure -- such as with several unpaired $d$ and/or $f$ electrons -- have to be addressed \cite{fleig_gasmcscf}. 
The present approach is also directly applicable to molecules.

The article is structured as follows: In section \ref{SEC:THEORY} the theory of the atomic electric quadrupole moment is
briefly reviewed and the present implementation using relativistic configuration interaction (CI) wavefunctions for
electronic ground and excited states is described. The same method is in the present also applied in the calculation of
the magnetic hyperfine interaction constant. Section \ref{SEC:RESULTS} contains the applications, first to the
beryllium (Be) atom as a test system for verification of the present implementation, then to the radium mono-cation
(Ra$^+$) as a more complex system where relativistic effects are strong. Finally, predictions are made for
EQMs of ground and excited states of the thulium (Tm) atom, in the case of excited states of Tm for the first time. In 
the final section \ref{SEC:CONCL} conclusions from the present findings are drawn.
\section{Theory}
\label{SEC:THEORY}
\subsection{Interaction energy}

An arbitrary charge distribution immersed in an external electric field ${\bf{E}}$ gives rise to an electrostatic 
interaction energy \cite{jackson}, the second-order term of which is written out as
\begin{equation}
        W_2 = -\frac{1}{6}\sum\limits_{i,j}\, Q_{ij}\,
                          \left. \frac{\partial E_j({\bf{x}})}{\partial x_i} \right|_{{\bf{x}}={\bf{x}}_0}
\end{equation}
where ${\bf{x}}_0$ is some conveniently chosen expansion point, ${\bf{Q}}$ is the rank-2 electric quadrupole
moment tensor of the charge distribution and $\frac{\partial E_j({\bf{x}})}{\partial x_i}$ is a component of
the electric-field gradient. In the present case the charge distribution is represented by the electron
shells of the atomic systems under consideration. The atomic nucleus is described by a spherical Gaussian charge
distribution and thus does not contribute to the EQM of the atom.

\subsection{Atomic electric quadrupole moment}

The relevant element of the rank-2 tensor $Q$ of the atomic electric quadrupole moment has been defined as
\cite{Sundholm_Olsen_QuaMom1993,Itano_QuaMom2006,ludlow_rmp_2015}, in \au,
\begin{equation}
	Q_{zz} := \left< \Psi \left| -\sum\limits_{i=1}^n\, \sqrt{\frac{4\pi}{5}}\; r^2(i)\, Y_{2,0}(\vartheta,\varphi)
	\right| \Psi \right>
	\label{EQ:AQM_DEF}
\end{equation}
with $\Psi$ the atomic wavefunction, $n$ the number of electrons, $r$ the radial electron coordinate and $Y_{\ell,m_{\ell}}$
a spherical harmonic. In Condon-Shortley convention we have
\begin{equation}
	Y_{2,0}(\vartheta,\varphi) = \frac{1}{4}\, \sqrt{\frac{5}{\pi}}\, \left( 3\cos^2(\vartheta) - 1 \right)
	\label{EQ:Y20_DEF}
\end{equation}
Therefore,
\begin{equation}
	Q_{zz} = -\frac{1}{2}\, \left< \Psi \left| \sum\limits_{i=1}^n\, r^2(i)\, \left( 3\cos^2(\vartheta) - 1 \right)\right| \Psi \right>
	\label{EQ:AQM_1}
\end{equation}
An elementary coordinate transformation yields
\begin{equation}
	r^2\, \left( 3\cos^2(\vartheta) - 1 \right) = 2z^2 - x^2 - y^2
	\label{EQ:COTR}
\end{equation}
in terms of cartesian coordinates, and so
\begin{eqnarray}
	\nonumber
	Q_{zz} &=& -\frac{1}{2}\, \left< \Psi \left| \sum\limits_{i=1}^n\, \left( 2z^2(i) - x^2(i) - y^2(i) \right) \right| \Psi \right> \\
	       &=& -\frac{1}{2}\, \left\{ \left< \Psi \left| \sum\limits_{i=1}^n\, 2z^2(i) \right| \Psi \right>
	                                  -\left< \Psi \left| \sum\limits_{i=1}^n\, x^2(i) \right| \Psi \right>
	                                  -\left< \Psi \left| \sum\limits_{i=1}^n\, y^2(i) \right| \Psi \right>
			          \right\}
\end{eqnarray}
The tensor element is evaluated by calculating the three resulting matrix elements over cartesian one-electron operators,
the ``second moments''.
In the present case, however, $\left|\Psi\right> \equiv \left| \alpha\, J\, M_J \right>$ is a relativistic atomic Dirac 
wavefunction where $\alpha$ represents quantum numbers other than the total angular momentum
$J$. 
Since a linear symmetry double group (in the present case $D_{32h}^*$ which is an abelian subgroup of $D_{\infty h}^*$)
is used instead of full rotational atomic symmetry the wavefunctions are obtained for individual $M_J$ states.
In practice, $J$ for a given eigenvector is inferred by calculating a sufficient 
number of degenerate eigenvectors in the different relevant $M_J$ subspaces. 
The individual states are represented by relativistic configuration interaction wavefunctions
\begin{equation}
        \left| \alpha\, J\, M_J \right> \equiv \sum\limits_{I=1}^{\rm{dim}{\cal{F}}^t(M,n)}\,
                                       c_{(\alpha, J, M_J),I}\, ({\cal{S}}{\overline{\cal{T}}})_I \rvac
        \label{EQ:AT_WF}
\end{equation}
where ${\cal{F}}^t(M,n)$ is the symmetry-restricted sector of Fock space with $n$ electrons in $M$ four-spinors,
${\cal{S}} = a^{\dagger}_i a^{\dagger}_j a^{\dagger}_k \ldots$ is a string of spinor creation operators,
${\overline{\cal{T}}} = a^{\dagger}_{\overline l} a^{\dagger}_{\overline m} a^{\dagger}_{\overline n} \ldots$
is a string of creation operators of time-reversal transformed spinors. The determinant expansion coefficients
$c_{(\alpha, J, M_J),I}$ are generally obtained as described in refs. \cite{fleig_gasci,fleig_gasci2}
by diagonalizing the Dirac-Coulomb Hamiltonian (in \au)
\begin{equation}
 \hat{H}^{\text{Dirac-Coulomb}}
      = \sum\limits^n_j\, \left[ c\, \boldsymbol{\alpha}_j \cdot {\bf{p}}_j + \beta_j c^2
    - \frac{Z}{r_j}{1\!\!1}_4 \right]
        + \sum\limits^n_{j,k>j}\, \frac{1}{r_{jk}}{1\!\!1}_4
 \label{EQ:DC_HAMILTONIAN}
\end{equation}
in the basis of the states $({\cal{S}}{\overline{\cal{T}}})_I \rvac$,
where the indices $j,k$ run over electrons, $Z$ is the proton number, and
$\boldsymbol{\alpha},\beta$ are standard Dirac matrices.
The framework for the present implementation is the relativistic electronic-structure program package DIRAC
\cite{DIRAC_JCP} where a locally modified version of the code is used.

In the present paper the electric quadrupole tensor component is evaluated for the microstate with $M_J = J$, \ie
\begin{equation}
        Q_{zz} = -\frac{1}{2}\, \left< \alpha\, J\, J \left| \sum\limits_{i=1}^n\, \left( 2z^2(i) - x^2(i) - y^2(i) \right) \right|  \alpha\, J\, J \right>.
\end{equation}
based on the evaluation of properties using relativistic CI wavefunctions as described in Refs. \cite{knecht_luciparII,knecht_thesis}.
From the EQM expectation value for a given 
eigenvector with defined $M_J$ the EQM for the other $M_J$-components of the associated state $J$ can be calculated {\it{via}}
the reduced matrix element (RME) evaluated through the adapted form of the Wigner-Eckhart theorem:
\begin{equation}
        {\text{RME}} = \left< \alpha\, J || \hat{Q}_{zz} || \alpha\, J \right> = 
        \frac{\left< \alpha\, J\, M_J | \hat{Q}_{zz} | \alpha\, J\, M_J \right>\, \sqrt{2 J + 1}}
                                                                   {\left< J\, 2\, M_J\, 0 | J\, 2\, J\, M_J \right>}
        \label{EQ:QZZ_RME}
\end{equation}
where $\hat{Q}_{zz} = -\frac{1}{2}\, \sum\limits_{i=1}^n\, \left[ 2z^2(i) - x^2(i) - y^2(i) \right]$
and $\alpha$ denotes quantum numbers other than those of total electronic angular momentum $J$.
The Clebsch-Gordan coefficients (CGC) in the denominator of Eq. (\ref{EQ:QZZ_RME}) are calculated according
to Wigner's (1959) general definition \cite{wigner1959} as given in reference \cite{weissbluth}:
{\footnotesize{
\begin{eqnarray}
        \nonumber
        \left< j_1 j_2 m_{j_1} m_{j_2} | j_1 j_2 j m_j \right> &=& \delta(m_j,m_{j_1}+m_{j_2})
           \sqrt{\frac{\left( j_1+j_2-j\right)!\left( j+j_1-j_2\right)!\left( j+j_2-j_1\right)!(2j+1)}{\left(j+j_1+j_2+1\right)!}} \\
           && \hspace{-3.7cm} \times \sum\limits_k\, \frac{(-1)^k \sqrt{(j_1+m_{j_1})!(j_1-m_{j_1})!(j_2+m_{j_2})!(j_2-m_{j_2})!(j+m_j)!(j-m_j)!}}{k! (j_1+j_2-j-k)!(j_1-m_{j_1}-k)!(j_2+m_{j_2}-k)!
                                                            (j-j_2+m_{j_1}+k)!(j-j_1-m_{j_2}+k)!}
        \label{EQ:CGC}
\end{eqnarray}
}}

\subsection{Magnetic Hyperfine Interaction}

The magnetic hyperfine interaction constant has been implemented in the present electronic-structure methods as described
in Refs. \cite{Fleig2014,Fleig_Skripnikov2020}. For $n$ electrons in the field of nucleus $K$ it is defined in \au\ as
\begin{equation}
 A(K) = -\frac{\mu_{K}[\mu_N]}{2cIm_p M_J}\,
 \left< J\; M_J \right| \sum\limits_{i=1}^n\, \left( \frac{\boldsymbol{\alpha}_i \times {\bf{r}}_{iK}}{r_{iK}^3}
 \right)_z \left| J\; M_J \right>
 \label{EQ:A_CONST}
\end{equation}
where $\mu_{K}$ is the nuclear magnetic moment, $\frac{1}{2cm_p}$ is the nuclear magneton in \au,  $m_p$ is the proton rest 
mass, $I$ is the nuclear spin quantum number, and ${\bf{r}}$ is the electron position operator.

\section{Applications and results}
\label{SEC:RESULTS}
\subsection{Be}
\label{SEC:BE}

\subsubsection{Technical details}
The Gaussian basis set for Be is cc-pV6Z with added diffuse functions from the set aug-cc-pV5Z \cite{EMSL-basis2019},
amounting to $\{17s,10p,6d,5f,4g,3h,1i\}$ uncontracted functions. 
Spinors are optimized for the closed-shell ground state ($1s^2 2s^2$). 
A cutoff energy of $100$ \au\ is used for the virtual spinor set. A Full CI expansion is used which includes the entire set
of Triple and Quadruple excitations, amounting to $\approx 29$ million Slater determinants.

\subsubsection{Results and discussion}
The above-defined model yields $Q_{zz} (2s^1 2p^1; {^3P_2}) = 2.2721$ \au for this Be excited state. For comparison, the
four-electron limit result from Ref. \cite{Sundholm_Olsen_QuaMom1993} is 
$Q_{zz} (2s^1 2p^1; {^3P_2}) = 2.265$ \au  This differs from the present result by only roughly $0.3$\%
confirming the reliability of the present implementation. The small difference is not explained by relativistic effects
which had been found to be smaller than $0.01$\% but rather by the fact that in Ref. \cite{Sundholm_Olsen_QuaMom1993}
a finite-element method is used whereas in the present case a finite Gaussian basis set is employed.

As a further consistency test $Q_{zz} (2s^1 2p^1; {^3P_2, M_J=1}) = -1.136$ \au\ is calculated explicitly from the
relevant $M_J$ eigenvector. The value for $Q_{zz} (2s^1 2p^1; {^3P_2, M_J=1})$ but calculated from 
$Q_{zz} (2s^1 2p^1; {^3P_2, M_J=2})$ and RME($2s^1 2p^1; {^3P_2}$) obtained from Eq. (\ref{EQ:QZZ_RME}) is indeed identical.

\subsection{Ra$^+$}
\label{SEC:RA}

\subsubsection{Technical details}

For Ra$^+$ a Gaussian basis set of quadruple-zeta quality is used where all $\{6s,6p,7s,7p\}$-correlating and all
dipole-polarizing primitive functions have been included \cite{dyall_s-basis}. 
The Dirac spinors are optimized by diagonalizing a Fock operator where a fractional occupation of $f = \frac{1}{12}$ per
$7s$ and $6d$ spinor is used. Effectively, this yields a Dirac-Hartree-Fock state that is
averaged over the {$^2S_{1/2}$} ground term and the {$^2D_{3/2,5/2}$} excited terms with spinors that are not biased towards 
any of the corresponding states. In correlated models the nine outermost electrons from shells $6s,6p,7s$ are considered.

\begin{table}[h]
        \caption{\label{TAB:QZZ_RA+} 
	   Atomic electric quadrupole moments and level energies for state $k$, defined as
           $\Delta\varepsilon(k) = \varepsilon(k) - \varepsilon({^2S}_{1/2})$ with $\varepsilon$ the total electronic energy,
	   for Ra$^+$
        }

 \vspace*{0.3cm}
 \hspace*{-1.3cm}
 \begin{tabular}{l|lcc}
	 Excited state         &  CI model               & $\Delta\varepsilon$ [cm$^{-1}$] &  $Q_{zz}$ [\au]  \\ \hline
	 ${^2D}_{3/2} (3d^1)$  &  DCHF ($7s^1$)                 & $15295$  &  $4.944$      \\
			       &  DCHF (av.)                    & $13053$  &  $3.312$      \\
			       &  SD9\_10au                     & $12171$  &  $2.998$      \\ 
			       &  SDT9\_10au                    & $12083$  &  $2.879$      \\ \hline
			       &  Final                         & {\bf{12083}}  &  {\bf{2.88(12)}} \\ \hline
			       &  Other theory                  &          &  $2.84(3)$\cite{Safronova_EQM_Ra+_2009}\\
			       &                                &          &  $2.90(2)$\cite{Sahoo_Ra+_2007}       \\ \hline
			       &  Exp. \cite{nist_raII_2022}    & $12084.3$&               \\ \hline
	 ${^2D}_{5/2} (3d^1)$  &  DCHF ($7s^1$)                 & $16021$  &  $7.309$      \\
			       &  DCHF (av.)                    & $13853$  &  $4.964$      \\
			       &  SD9\_10au                     & $13639$  &  $4.559$      \\ 
			       &  SDT9\_10au                    & $13654$  &  $4.402$      \\ \hline
			       &  Final                         & {\bf{13654}}  &  {\bf{4.40(16)}} \\ \hline
			       &  Other Theory                  &          &  $4.34(4)$\cite{Safronova_EQM_Ra+_2009} \\ 
			       &                                &          &  $4.45(9)$\cite{Sahoo_Ra+_2007} \\ \hline
			       &  Exp. \cite{nist_raII_2022}    & $13743.0$&               \\ \hline
  \end{tabular}
\end{table}

\subsubsection{Results and discussion}
\label{SEC:RA_RES}

EQMs and level energies for Ra$^+$ are compiled in Table \ref{TAB:QZZ_RA+}. The DCHF results using spinors specific to the
electronic ground state of Ra$^+$ ($7s^1$ configuration) exhibit large deviations from experiment and from reliable theoretical
results. The spinor averaging (DCHF (av.)) rectifies this problem to a large degree. Including electron correlation effects at
up to the level of Double excitations (model SD9\_10au) based on the state-averaged spinors diminishes $Q_{zz}({^2D}_{3/2})$
by roughly $10$\% and $Q_{zz}({^2D}_{5/2})$ by $8$\%, respectively. Triple excitations are also not unimportant, further 
quenching $Q_{zz}({^2D}_{3/2})$ and $Q_{zz}({^2D}_{5/2})$ by $3-4$\%.
The term energies for both excited states are in excellent agreement with experiment in the most accurate model, SDT9\_10au,
where deviations are less than $1$\%. Present uncertainties are estimated conservatively to be at most as large as the effect 
of Triple excitations, accounting for excitation ranks higher than Triples and basis-set truncations.

The present final results fall in between those from Refs. \cite{Safronova_EQM_Ra+_2009} and \cite{Sahoo_Ra+_2007} and are,
considering uncertainties, compatible with both. However, from the present series of calculations, it can be concluded that
Quadruple excitations and beyond will lead to a further -- albeit small -- downward correction to $Q_{zz}$, likely on the
order of $1-2$\%. A further downward correction of a few percent is to be expected from the inclusion of basis functions with 
high angular momentum ($\ell > 4$), as has been pointed out in Refs. \cite{Safronova_EQM_2008,Safronova_EQM_Ra+_2009}. The
present basis set terminates at $\ell = 4$ for the large-component of the Dirac wavefunctions.
Thus, the present EQM results for Ra$^+$ are in very good agreement with those from Ref. \cite{Safronova_EQM_Ra+_2009} which 
remain the most accurate to date. This is a further confirmation of the reliability of the present approach to calculating
atomic EQMs, which was the goal for this study on Ra$^+$.

\subsection{Tm}
\label{SEC:TM}

The afore-going sections have set the stage for making reliable predictions for EQMs in systems where they are so far unknown.

\subsubsection{Technical details}

Two Gaussian basis sets are employed for Tm, Dyall's ccpVTZ set with all $4f,6s,5s,5p,5d$-correlating and $4f$ 
dipole-polarizing functions added amounting to $\{30s,24p,18d,13f,4g,2h\}$ functions and Dyall's ccpVQZ set including all 
valence-correlating and the $4f$ dipole-polarizing functions \cite{dyall_4f}, amounting to $\{35s,30p,19d,16f,6g,4h,2i\}$ functions.
The Dirac spinors are optimized by diagonalizing a Fock operator where a fractional occupation of $f = \frac{13}{14}$ per
spinor with $\ell=3$ is used ($4f^{13}\; 6s^2$ ground configuration). Effectively, this yields a Dirac-Hartree-Fock state that is 
averaged over the {$^2F_{7/2}$} ground term and the {$^2F_{5/2}$} excited term and spinors that are not biased towards any 
of the corresponding states.

\begin{table}[h]
        \caption{\label{TAB:QZZ_TM_TZ} 
	   Atomic electric quadrupole moments for Tm ${^2F}_{7/2} (4f^{13}\, 6s^2)$ using various CI models, TZ basis
        }

 \vspace*{0.3cm}
 \hspace*{-1.3cm}
 \begin{tabular}{lr}
           CI model (number of virtual functions) &  $Q_{zz}$ [\au] \\ \hline
	   DCHF                                   &  $-0.2711$    \\
 	   SD15\_0.2au $(1s,1d;2p)$               &  $ 0.1128$    \\ 
 	   SD15\_1au $(2s,3d,1g;3p,2f)$           &  $-0.0101$    \\ 
 	   SD15\_2au $(3s,4d,2g;4p,3f)$           &  $-0.1170$    \\ 
 	   SD15\_4au $(3s,4d,2g;4p,4f,1h)$        &  $-0.1920$    \\ 
	   SD15\_6au $(4s,5d,2g;5p,5f,1h)$        &  $-0.2319$    \\
	   SD15\_10au $(4s,6d,3g;5p,6f,1h)$       &  $-0.2519$    \\ 
	   SD15\_20au $(5s,7d,3g;6p,7f,1h)$       &  $-0.2543$    \\ 
	   SD15\_50au $(6s,8d,4g;7p,8f,2h)$       &  $-0.2530$    \\
	   SD15\_130au $(7s,9d,4g;8p,9f,2h)$      &  $-0.2531$    \\ \hline
	   SD23\_10au                             &  $-0.2470$    \\ \hline
	   SD33\_10au                             &  $-0.2473$    \\ \hline
 	   SDT15\_0.2au                           &  $ 0.1181$    \\ 
 	   SDT15\_1au                             &  $ 0.0509$    \\ 
 	   SDT15\_2au                             &  $-0.0139$    \\ 
 	   SDT15\_4au                             &  $-0.0498$    \\ 
 	   SDT15\_6au                             &  $-0.0472$    \\ 
 	   SDT15\_10au                            &  $-0.0299$    \\ 
 	   SDT15\_20au                            &  $-0.0311$    \\ 
 	   SDT15\_50au                            &  $-0.0304$    \\ \hline
 	   SDTQ15\_2au                            &  $ 0.021 $    \\ 
 	   SDTQ15\_4au                            &  $ 0.013 $    \\ \hline
	   SDTQ15\_10au (est.)                    &  $ 0.067 $    \\ \hline
  \end{tabular}
\end{table}

\subsubsection{Results and discussion}
Tables \ref{TAB:QZZ_TM_TZ} and \ref{TAB:QZZ_TM_QZ} show results for the EQM of the electronic ground term ${^2F}_J$.
Valence correlation effects from up to Double excitations are converged at the percent level when the virtual spinor space
is truncated at $20$ \au  These effects decrease $Q_{zz}$ -- on the absolute -- by about $6$\%. However, when full Triple
excitations are introduced in addition, the EQM is quenched by an astonishing order of magnitude. This strong quenching is
basis-set dependent to only about $6$\% which justifies a deeper investigation using the smaller TZ basis set only.

The strongest positive contribution to $Q_{zz}$ from excitations into the virtual spinor space is observed at the very low 
cutoff of $0.2$ \au\  Further cutting down on this virtual set reveals that the positive contribution is mainly due to
Double excitations of the type $6s^2 \rightarrow 5d_{5/2,5/2}^2$, the amplitude (CI coefficient) of which diminishes
strongly as the dimension of the virtual space is increased. It is remarkable that the residual quenching of $Q_{zz}$
is about $+0.24$ \au\ for the model SDT15\_20au which leads to a value for $Q_{zz}$ that is roughly one order of
magnitude smaller on the absolute than the DCHF value. Expressed in different terms, we here observe an electron
correlation effect (difference between a given CI model and Hartree-Fock theory) of nearly 90\% ! This
extraordinary situation can be attributed to the fact that the open-$f$-shell contribution to $Q_{zz}$ of the thulium
atom is rather small and that even a small amplitude on the contributions that arise from $d$-shell occupations which, 
furthermore, have opposite sign, can nearly cancel the latter.

The discussed cancellation leads to values for $Q_{zz}$ that are close to zero and thus manifestly very difficult to 
describe accurately, i.e., with small relative errors. The inclusion of excitation ranks higher than full Triples is 
explored using the smaller (TZ) basis set, see Table \ref{TAB:QZZ_TM_TZ}. Due to the extreme computational demand a 
truncation of the virtual space has to serve as a further approximation. 
When this truncation is set to $2$ \au\ the difference between the models 
SDT15 and SDTQ15 is $+0.034$ \au\ However, with this spinor space the model SDT15 is qualitatively incorrect. A
truncation at $4$ \au\ yields an SDT15 result that agrees qualitatively with the converged result from SDT15\_50au.
The corresponding expansion for SDTQ15\_4au comprises roughly $6$ billion expansion terms, close to the limits of
computational feasibility with the current code.

\begin{figure}[h]
        \caption{\label{FIG:QZZ_TM_TZ} 
	   EQM (\au) of ${^2F}_{7/2}$ Tm ground state using various CI models and virtual cutoffs, TZ basis; 
	   horizontal lines display converged values for the respective model. The convergence patterns for the two
	   largest calculations are shown explicitly.
        }

 \vspace*{0.0cm}
 \hspace*{-1.5cm}
	\includegraphics[width=18.0cm,angle=0.0]{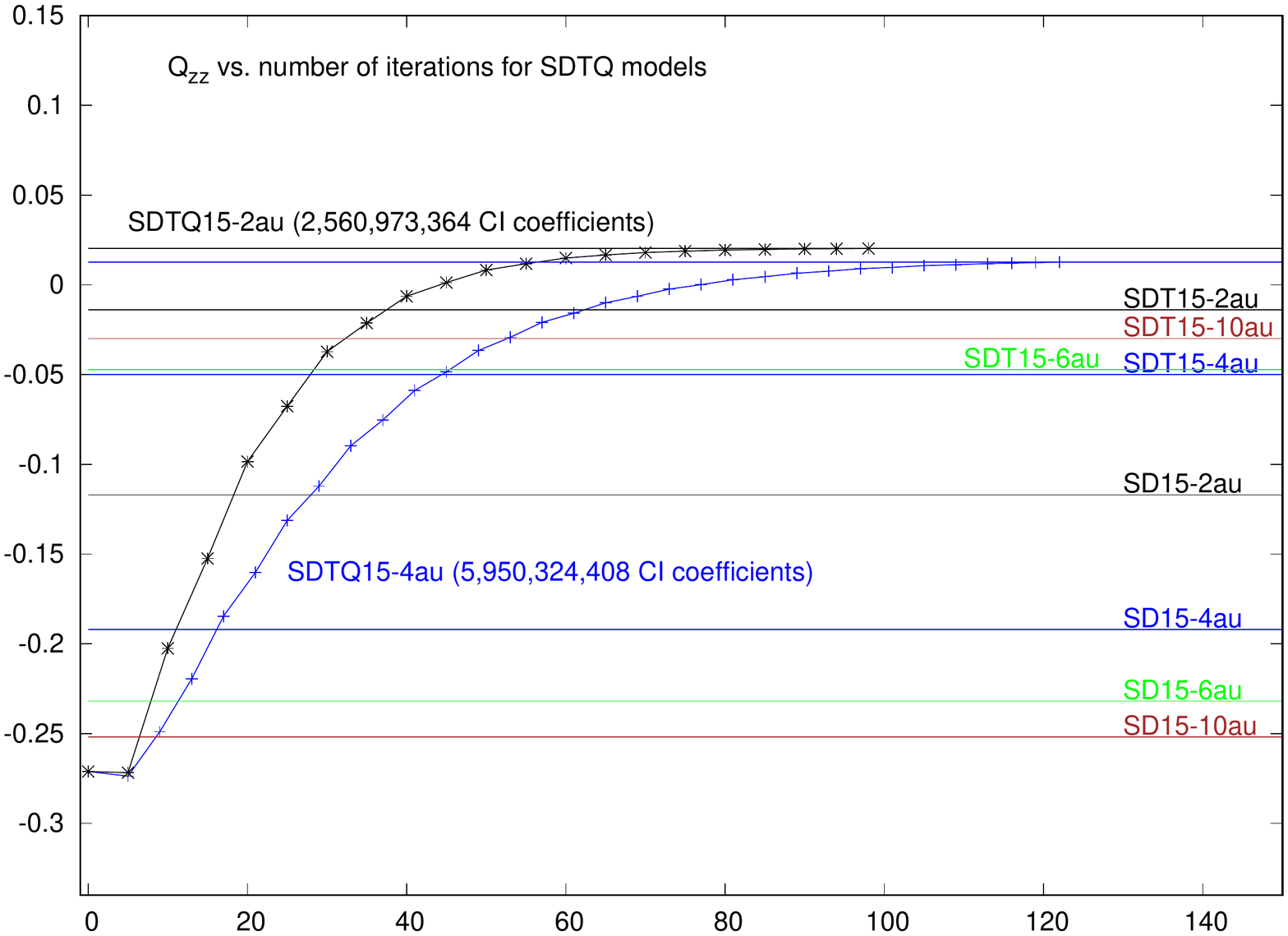}
\end{figure}

A correction due to full Quadruple excitations -- which is not negligible in the present case -- is thus obtained as follows. 
The base value is provided by the model
SDT15\_10au. Since correlation contributions at any excitation rank are nearly converged at a 
virtual cutoff of $10$ \au\ but this spinor space creates a configuration space too large to be treated explicitly when Q
excitations are taken into account, the result for the model SDTQ15\_10au is estimated by the following formula:
\begin{eqnarray*}
	Q_{zz}({\rm{SDTQ15\_10au}}) &:=& Q_{zz}({\rm{SDT15\_10au}}) + \Delta Q[Q_{zz}({\rm{10au}})]
\end{eqnarray*}
where
\begin{eqnarray*}
	\Delta Q[Q_{zz}({\rm{10au}})] &:=& \left(Q_{zz}({\rm{SDTQ15\_4au}})-Q_{zz}({\rm{SDT15\_4au}})\right) \\
				  && \times \frac{\left(Q_{zz}({\rm{SDT15\_10au}})-Q_{zz}({\rm{SD15\_10au}})\right)}
				              {\left(Q_{zz}({\rm{SDT15\_4au}})-Q_{zz}({\rm{SD15\_4au}})\right)}
\end{eqnarray*}
This scales the correction due to Quadruple excitations by the ratio of the Triples correction for different virtual cutoff 
values. In this estimation the assumption is made that the augmentation of the virtual spinor space affects
the Triples correction $\Delta T$ and the Quadruples correction $\Delta Q$ in an equivalent manner.
This way of obtaining the correction is also supported by the graphical representation in Fig. (\ref{FIG:QZZ_TM_TZ}).
$\Delta T$ is always positive (as is $\Delta Q$) and increases monotonously as a function of virtual cutoff (assumed
for $\Delta Q$). Since the correction due to higher excitation ranks is more than halved when going from $\Delta T$ to
$\Delta Q$ a correction $\Delta 5$ due to full quintuples (or to even higher excitation ranks) is not expected to
surpass $+0.05$ \au\ for $Q_{zz}({^2F}_{7/2})$.

Possible corrections due to the use of the larger QZ basis set are investigated through the calculations presented in Table
\ref{TAB:QZZ_TM_QZ}.
Interestingly, the most elaborate comparable model SDT15\_50au does not yield a significant basis-set correction ($\Delta$
QZ $\approx -0.0001$ \au). Furthermore, a correction due to core-valence correlation by including single-hole configurations
(model S8\_SD23\_10au) and in addition double-hole configurations (model SD23\_10au) in the Tm $5s,5p$ shells also result in
negligibly small corrections. The same is true for correlations due to single- and double-hole configurations in the Tm $4d$
shell (see Table \ref{TAB:QZZ_TM_TZ}).
The inclusion of additional functions of very high angular momentum ($h$ ($\ell=5$) and $i$ ($\ell=6$))
in the basis set does not affect the EQM of the Tm atom appreciably, as can be seen by comparing the models SDT15\_50au in
Tables \ref{TAB:QZZ_TM_TZ} and \ref{TAB:QZZ_TM_QZ}. Also, the Triples correction $\Delta T$ increases by only about $2.5$\%
when going from the TZ to the QZ basis, a further indication that many-body effects are well described using the TZ basis set.

The present final best result is, therefore, ${\boldsymbol{Q_{zz}({^2F}_{7/2}) \approx 0.07}}$ \au  I assign to this
a rather conservative uncertainty of $100$\%, almost entirely due to the neglect of higher CI excitation ranks in even the 
most highly correlated atomic wavefunction (SDTQ). Given the small value of the EQM this uncertainty translates into only
$0.07$ \au

\begin{table}[h]
        \caption{\label{TAB:QZZ_TM_QZ} 
	   Atomic electric quadrupole moments and level energies for state $k$, defined as
           $\Delta\varepsilon(k) = \varepsilon(k) - \varepsilon({^2F}_{7/2})$ with $\varepsilon$ the total electronic energy,
	   for Tm using various CI models, QZ basis
        }

 \vspace*{0.3cm}
 \hspace*{-1.3cm}
 \begin{tabular}{l|lcc}
		 State                   &  CI model (virtual functions)      & $\Delta\varepsilon$ [cm$^{-1}$] &  $Q_{zz}$ [\au] \\ \hline
	 ${^2F}_{7/2} (4f^{13}\, 6s^2)$  &  DCHF                                 & $    0$     &  $-0.2711$    \\
					 &  SD15\_10au $(6s,5d,4g,1i;7p,7f,2h)$  & $    0$     &  $-0.2531$    \\ 
					 &  SD15\_20au $(7s,6d,5g,1i;8p,8f,3h)$  & $    0$     &  $-0.2546$    \\ 
					 &  SD15\_50au $(8s,7d,5g,2i;9p,10f,3h)$ & $    0$     &  $-0.2556$    \\ 
					 &  SDT15\_10au                          & $    0$     &  $-0.0349$   \\ 
					 &  SDT15\_20au                          & $    0$     &  $-0.0292$    \\ 
					 &  SDT15\_50au                          & $    0$     &  $-0.0305$    \\ 
					 &  S8\_SD23\_10au                       & $    0$     &  $-0.2511$    \\
					 &  SD23\_10au                           & $    0$     &  $-0.2534$    \\ \hline
					 &  Exp. \cite{nist_raII_2022} & $   0$     &               \\ \hline
	 ${^2F}_{5/2} (4f^{13}\, 6s^2)$  &  DCHF                                 & $ 9016$     &  $-0.2203$    \\
                                    	 &  SD15\_5au                            & $ 9277$     &  $-0.1863$    \\
	                                 &  SD15\_10au                           & $ 9215$     &  $-0.2059$    \\
	                                 &  SD15\_20au                           & $ 9143$     &  $-0.2074$    \\
					 &  SDT15\_20au                          & $ 9028$     &  $-0.0257$    \\ 
	                                 &  SD15\_50au                           & $ 9128$     &  $-0.2083$    \\ 
					 &  S8\_SD23\_10au                       & $ 9199$     &  $-0.2043$    \\
					 &  SD23\_10au                           & $ 9158$     &  $-0.2061$    \\ \hline
					 &  Exp. \cite{nist_raII_2022} & $ 8771.243$ &               \\ \hline
  \end{tabular}
\end{table}

As can be inferred from the above detailed discussion and the results in Table \ref{TAB:QZZ_TM_QZ} it can be said with certainty 
that $0 < Q_{zz}({^2F}_{5/2}) < Q_{zz}({^2F}_{7/2})$. Thus, the EQM of the electronically excited clock state also has near-zero 
EQM.

The result by Sukachev et al. \cite{PRA_Tm_2016} obtained with the \verb+COWAN+ code of $Q_{zz}({^2F}_{7/2}) \approx 0.5$ \au\
differs from the present final result by roughly $0.4$ \au\ This is a large relative and also a significant absolute difference, 
given the observed discrepancies from different electronic-structure models as compared for the Ra$^+$ ion in section 
\ref{SEC:RA_RES} at the highest level of accuracy. The spread of final values for Ra$^+$ is about an order of magnitude smaller 
than $0.4$ \au
Since Sukachev et al. do not give any details of their calculation it is not possible to analyze the discrepancy for Tm.

\subsubsection{Hyperfine interaction}

However, a qualitative judgement on the present EQM results is possible in an indirect manner. Since closed-shell contributions 
to the EQM are zero the spin density in the atomic state is linked to the EQM. The same is true for the magnetic hyperfine
interaction constant. Thus, the correlated wavefunctions presently optimized for the description of the Tm EQM are expected
to also describe the corresponding hyperfine interaction constant correctly. 

\begin{table}[h]
        \caption{\label{TAB:A_TM} 
	   Magnetic hyperfine interaction constant $A$ for {$^{169}$Tm} ${^2F}_{7/2} (4f^{13}\, 6s^2)$ 
	   calculated with Eq. (\ref{EQ:A_CONST}) using various CI models and basis sets
        }

 \vspace*{0.3cm}
 \hspace*{-1.3cm}
 \begin{tabular}{l|l}
	  CI model/basis set                      & $A$ [MHz]        \\ \hline
	  SDT15\_2au/TZ                           & $-388.5$         \\
	  SDT15\_4au/TZ                           & $-391.0$         \\
	  SDT15\_6au/TZ                           & $-396.8$         \\
	  SDT15\_10au/TZ                          & $-397.5$         \\
	  SDT15\_20au/TZ                          & $-399.9$         \\
	  SDT15\_50au/TZ                          & $-399.6$         \\
	  SDTQ15\_2au/TZ                          & $-388.4$         \\
	  SDTQ15\_4au/TZ                          & $-390.8$         \\ \hline
	  SDT15\_10au/QZ                          & $-397.2$         \\
	  SDT15\_20au/QZ                          & $-398.2$         \\
	  SDT15\_50au/QZ                          & $-399.5$         \\ \hline
	  Exp.\cite{giglberger_Tm_HypFin_1966}    & $-374.137661(3)$
 \end{tabular}
\end{table}

Table \ref{TAB:A_TM} lists the results for $A$ using wavefunctions from the most accurate models for calculating the EQM.
The spin density in the correlated wavefunction of the Tm ${^2F}_{7/2}$ ground state largely resides in $f$ (and $d$) states 
which explains why $A$ is comparatively small.
Also here the basis set effect is negligibly small. Higher excitations than Triples do not affect $A$ substantially. The
result with the largest deviation from the experimental value for $A$ is obtained with the model SDT15\_50au/QZ which 
differs from the experimental value by only about $6.8$\%. This finding confirms the quality of the wavefunctions used for
calculating the EQM in the present work.

\section{Conclusions and Outlook}
\label{SEC:CONCL}
In the present paper an accurate method for the calculation of atomic electronic electric quadrupole
moments is presented. The applications to the Be atom and to the Ra$^+$ ion demonstrate the reliability
of the method.

The quadrupole shift has been identified as one of the contributors to the uncertainty budget for a Tm
optical clock using its ground-state fine-structure components \cite{PRA_Tm_2016}, albeit not the leading one at
the moment.
An elaborate, detailed, and high-level study on the Tm atom in the present work shows that the EQM in its 
${^2F}_{7/2} (4f^{13}\, 6s^2)$ ground state is exceptionally small.
Thus, the application of methods for systematic cancellation or suppression of the quadrupole shift 
\cite{Oskay_QuaSupp2005,PhysRevLett.95.033001} may not be necessary in a thulium optical clock.

On the methodological side the present work is motivated by the general importance of electric multipole
moments and electric transition multipole moments in many areas of atomic and molecular physics. The
author has also implemented E1 and E2 transition moments into the present relativistic correlated
many-body methods, applications of which will be the subject of publications in the very near future.  
E1 transition moments allow for the calculation of atomic dispersion coefficients which contribute to a 
clock's uncertainty through the van der Waals interaction. E2 transition moments are of importance for
instance in the field of parity nonconservation \cite{Fortson_PRL1993}.
\clearpage

\bibliographystyle{unsrt}
\newcommand{\Aa}[0]{Aa}

\clearpage

\end{document}